\documentclass{article}
\usepackage{spconf,amsmath,graphicx}
\usepackage{hhline}
\usepackage{comment}
\usepackage{color}
\usepackage{hyperref}

\ninept
\title{Music Enhancement via Image Translation and Vocoding}
\twoauthors
  {Nikhil Kandpal\sthanks{Work done during an internship with Adobe Research}}
	{University of North Carolina at Chapel Hill\\
	Computer Science Department\\
	Chapel Hill, NC, USA}
  {Oriol Nieto, Zeyu Jin}
	{Adobe Research\\
	San Francisco, CA, USA}

\begin{document}
%
\maketitle

\begin{abstract}
Consumer-grade music recordings such as those captured by mobile devices typically contain distortions in the form of background noise, reverb, and microphone-induced EQ. This paper presents a deep learning approach to enhance low-quality music recordings by combining (i) an image-to-image translation model for manipulating audio in its mel-spectrogram representation and (ii) a music vocoding model for mapping synthetically generated mel-spectrograms to perceptually realistic waveforms. We find that this approach to music enhancement outperforms baselines which use classical methods for mel-spectrogram inversion and an end-to-end approach directly mapping noisy waveforms to clean waveforms. Additionally, in evaluating the proposed method with a listening test, we analyze the reliability of common audio enhancement evaluation metrics when used in the music domain.
\end{abstract}
\section{Introduction}
\label{sec:intro}

With the rise of Internet influencers and music hobbyists, a large portion of music content is created with cheap and accessible recording devices in non-treated environments. While being audible, these recordings often have degraded quality stemming from background noise, unpleasant reverb, and resonance caused by the microphone and the environment. This prompts us to investigate quality enhancement for music signals, transforming low-quality amateur recordings into professional ones.

The main difficulty of such an endeavor is that so many aspects of the low-quality recording setup are unknown. Parameters of the recording device, such as frequency response characteristics, vary drastically across different hardware. Additionally, acoustic properties such as the size, shape, and reflectivity of the recording environment vary between different recording setups. 
Finally, background noise is hard to capture and generalize, especially non-stationary noise. A solution that faithfully transforms a low-quality recording into what it would sound like recorded professionally must implicitly or explicitly infer all of these aspects from the signal alone. In speech enhancement, end-to-end methods such as HiFi-GAN~\cite{su2020hifigan} and Demucs~\cite{demucs2020} achieve this by extracting the speech source from a mixture of sources. 
However, music signals are often polyphonic, i.e., there can be an arbitrary number of sources to be extracted at once. 
Moreover, the perception of music quality typically differs from that of speech. 
For example, human listeners may find reverb pleasant in music, while it is usually undesired in speech. 
Therefore, we aim to develop a solution that works for polyphonic signal enhancement and reflects the unique qualities of music perception. 

Our approach performs enhancement on the recording's mel-spectrogram representation. This is achieved by treating the mel-spectrogram as an image and training an image-to-image translation model similar to Pix2Pix \cite{isola2018imagetoimage} to transform a low-quality mel-spectrogram into that of a high-quality signal. 
We hypothesize that it is easier to enhance polyphonic signals in the mel-spectrogram domain as polyphonic sources are additive and have a very small temporal span compared to waveforms.  Finally, to map generated high-quality mel-spectrograms to perceptually realistic waveforms, we train a vocoding model based on DiffWave~\cite{kong2021diffwave}. 
Training this model on only the high quality samples of music performance makes it robust to the artifacts that reside in the synthetic mel-spectrogram. 

We evaluate our approach by performing a listening test with 211 participants, and we show that this approach achieves a much better perceptual enhancement than several state-of-the-art techniques.
We also compare the subjective listening test scores with widely used audio quality metrics and suggest that, similar to speech enhancement, these metrics correlate poorly with human perception~\cite{su2020hifigan, su2021bandwidth}. 
With this work, we hope to motivate both future research in music enhancement as well as music quality perceptual metrics akin to those in the speech literature~\cite{manocha2021cdpam, reddy2021dnsmos}. To promote further research, audio samples generated in our experiments and source code are provided at our project website\footnote{\url{https://nkandpa2.github.io/music-enhancement}}.

In this paper, we refer to Pix2Pix models operating on mel-spectrograms as \textit{Mel2Mel} models and vocoding applied to the music domain as \textit{musecoding}. We summarize our contributions as follows:
\begin{itemize}
    \item A music enhancement model leveraging recent work on conditional image synthesis and vocoding.
    \item A generative process for simulating realistic low-quality music recordings from professional-quality recordings.
    \item An analysis of the reliability of common audio enhancement evaluation metrics in the music domain.
\end{itemize}

\begin{figure*}
\centering
  \includegraphics[width=0.9\textwidth]{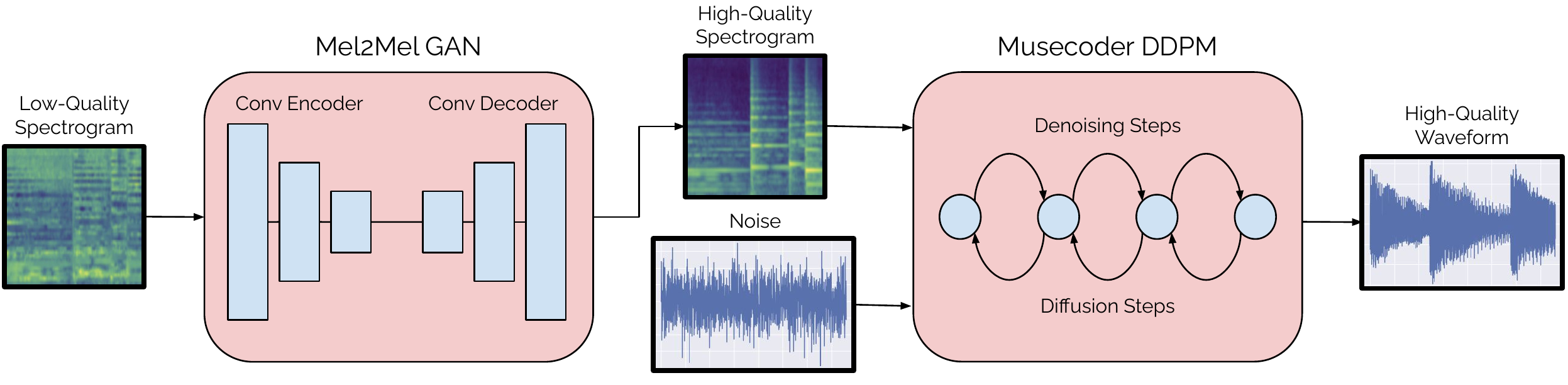}
  \caption{Model architecture of our Mel2Mel + Diffwave model. First, a low-quality mel-spectrogram is enhanced by a conditional GAN. The resulting synthetic mel-spectrogram is then ``musecoded'' into a waveform by a Denoising Diffusion Probabilistic Model (DDPM).}
  \label{fig:arch}
\end{figure*}

\section{Related Work}
\label{sec:related_work}
To our knowledge there is little prior work studying music quality enhancement. The work most similar to our contributions focuses on speech enhancement, conditional speech synthesis, or music source separation.

Early approaches to speech enhancement have used classical signal processing techniques such as Wiener filtering~\cite{543199} and non-negative matrix factorization~\cite{8462080}. More recently, deep learning-based methods have achieved state-of-the-art on speech enhancement. These methods either manipulate the audio in its magnitude spectrogram representation (followed by a spectrogram inversion method to recreate the corresponding waveform) \cite{7067387,7953226,Michelsanti_2017} or map directly from the low-quality waveform to a cleaned waveform \cite{pascual2019generalized,demucs2020,su2020hifigan}. 
Methods that operate on the time-frequency domain generally produce audible artifacts due to the use of phase reconstruction algorithms like the Griffin-Lim algorithm~\cite{1164317}. 
A recent work addresses this with neural-network based vocoders~\cite{polyak2021high}, yet its quality is not on par with an end-to-end approach~\cite{hifigan2}. 
Alternatively, methods that work on the time domain typically require more training steps~\cite{su2020hifigan}.

Conditional speech synthesis techniques produce speech waveforms from conditioning information such as magnitude spectrograms, a problem commonly known as vocoding. 
Some state-of-the-art vocoding methods involve using generative adversarial networks \cite{kumar2019melgan,you2021gan} or denoising diffusion probabilistic models \cite{kong2021diffwave,chen2020wavegrad} for generating audio. 

Music source separation focuses on taking a mix of multiple music ``stems'' (vocals, drums, etc.) and separating the mix into its individual sources. Some approaches to music source separation operate by masking spectrograms~\cite{article} or directly mapping the mix waveform to individual source waveforms~\cite{defossez2021music,Luo_2019}. 
The music enhancement problem is different than music source separation, since our goal is not only to extract all musical sources from a noisy mixture but also to reduce reverb and adjust EQ such that the listening experience is improved. 

\section{Methods}
\label{sec:methods}

\subsection{Modeling Approach}
\label{ssec:modeling_approach}
In this paper, we investigate the approach of enhancing music in its mel-spectrogram domain, as it is easier to represent complex harmonic structures and polyphonic sound sources. 
We then transform the resulting mel-spectrograms to waveforms through a Diffwave-based vocoder (a process that in this context could be more aptly named ``musecoding''). 
Decoupling waveform generation from mel-spectrogram enhancement allows us to train a musecoder that is not only robust to noise and other artifacts, but can also be used for any generation and enhancement task without the need of retraining. 
Figure~\ref{fig:arch} depicts a block diagram of our proposed architecture.
This approach is motivated by recent advances in vocoding that generate natural-sounding speech from mel-spectrograms~\cite{kong2021diffwave}.  

\subsection{Data Simulation}
\label{ssec:data_simulation}
The modeling techniques we consider in this paper require aligned pairs of high- and low-quality music recordings. 
To construct such a dataset, we assume access to high-quality recordings and define a generative process for simulating low-quality ones. 
First, we simulate the reverb and varied microphone placements of a non-professional recording environment by convolving the high-quality music signal with a room impulse response. Next, we apply additive background noise scaled to achieve a randomly sampled SNR between 5 and 30 dB. Finally, we simulate a low-quality microphone frequency response by applying 4-band equalization with randomly sampled gains between -15 and 15 dB and frequency bands from 0-200, 200-1000, 1000-4000, and 4000-8000 Hz.

\subsection{Mel-Spectrogram Enhancement with Mel2Mel}
\label{ssec:conditional_image_synthesis}
Our first step in music enhancement is modeling the distribution of high-quality mel-spectrograms conditioned on their low-quality counterparts. To estimate this distribution, we use existing work on image-to-image translation with conditional adversarial networks \cite{isola2018imagetoimage} in an approach similar to \cite{Michelsanti_2017}. 

In this framework a generator and a discriminator are trained using an aligned dataset of low and high-quality recording pairs. The generator maps from low to high-quality mel-spectrograms with the  objective of maximizing the discriminator's loss and minimizing the $\ell_1$ distance between the generated mel-spectrogram and the ground truth high-quality mel-spectrogram. 
The discriminator is trained to classify whether a given mel-spectrogram is generated or comes from the true data distribution. It performs this classification on a patch-wise basis, predicting a class for each patch in the input mel-spectrogram. For this reason, the discriminator acts as a learned loss function for the generator which enforces realistic local features and the $\ell_1$ loss enforces global consistency with the ground truth mel-spectrogram.

\subsection{Musecoding}
\label{ssec:music_vocoding}
Recent work has shown that deep learning models can generate perceptually realistic waveforms from speech mel-spectrograms. In our experiments, we evaluate the Diffwave \cite{kong2021diffwave} vocoder applied to music, a process that we call ``musecoding''.

Diffwave is a denoising diffusion probabilistic model (DDPM). This class of models defines a forward diffusion process which iteratively adds gaussian noise to audio waveforms from the training dataset. A model is then trained to estimate the reverse transition distributions of each noising step conditioned on the mel-spectrogram of the clean audio. Sampling from this model requires sampling noise from a standard gaussian and iteratively denoising using the reverse transition probability distributions from the model. For further discussion of DDPMs see \cite{kong2021diffwave} and \cite{ho2020denoising}.

As a musecoding baseline, we also consider mel-spectrogram inversion with inverse mel-scaling and the Griffin-Lim algorithm \cite{1164317}.

\section{Experiment Setup}
\label{sec:experimental_setup}

\subsection{Dataset}
We train and evaluate models on the Medley-solos-DB dataset \cite{lostanlen2017deep}, containing 21,572 three-second, single-instrument samples recorded in professional studios. We exclude the distorted electric guitar samples to avoid fitting our models to production effects. We use 5841 samples for training, 3494 for validation and the rest for testing. We start by downsampling our data to 16 kHz following the setup of prior vocoding work \cite{kong2021diffwave,kumar2019melgan}. 
This sample rate has shown to be favored by most speech enhancement work~\cite{su2020hifigan, demucs2020} and can be potentially super-resolved to 48 kHz with bandwidth extension techniques~\cite{su2021bandwidth}. Using the procedure described in Section \ref{ssec:data_simulation}, we generate a dataset of high- and low-quality recording pairs. For simulation of low-quality recordings, we source room impulse responses from the DNS Challenge dataset \cite{reddy2021interspeech} and realistic background noise from the ACE Challenge dataset \cite{7336912}. As a final step, we apply a low-cut filter to remove nearly inaudible low frequencies below 35 Hz and normalize the waveforms to have a maximum absolute value of 0.95. We find that this treatment helps improve our models' training stability. When evaluating, we apply the same treatment (low-cut filter at 35 Hz and normalization) before applying our enhancement models.

\begin{table}[]
    \centering
    \begin{tabular}{c|c}
         Model & MOS \textuparrow \\ \hhline{=|=}
         Clean & $4.39 \pm 0.05$\\ \hline
         Mel2Mel + Diffwave & $\mathbf{4.06 \pm 0.06}$\\ 
         Mel2Mel + Griffin-Lim & $3.01 \pm 0.09$ \\ 
         No Enhancement & $2.85 \pm 0.09$ \\

    \end{tabular}
    \caption{Mean Opinion Scores in a human listening test.}
    \label{tab:mos}
\end{table}

\subsection{Model Architectures and Hyperparameters}
In all experiments, we compute mel-spectrograms with 128 mel bins, an FFT size of 1024, and a 256 sample hop length. 
When training models that generate or are conditioned on mel-spectrograms, we use log-scale amplitudes to reduce the range of values and to avoid positive restrictions on our models' domain or range.

\begin{table*}[]
    \centering
    \begin{tabular}{c||c|c|c|c|c|c|c}
         Model & Random EQ & SNR 5 & SNR 10 & SNR 15 & DRR 0 & DRR 3 & DRR 6\\ \hhline{=|=|=|=|=|=|=|=}
         Clean & $4.35 \pm 0.06$ & $4.24 \pm 0.07$ & $4.27 \pm 0.06$ & $4.46 \pm 0.06$ & $4.28 \pm 0.04$ & $4.19 \pm 0.06$ & $4.42 \pm 0.06$\\  \hline
         Mel2Mel + Diffwave & $\mathbf{4.15 \pm 0.07}$  & $\mathbf{4.01 \pm 0.08}$ & $\mathbf{4.24 \pm 0.06}$ & $\mathbf{3.96 \pm 0.09}$ & $3.77 \pm 0.06$ & $ 3.84 \pm 0.06 $ & $3.96 \pm 0.08$\\ 
         Mel2Mel + Griffin-Lim & $2.98 \pm 0.1$ & $3.10 \pm 0.08$  & $3.53 \pm 0.09$ & $3.18 \pm 0.11$ & $2.82 \pm 0.07$ & $2.77 \pm 0.09$ & $2.99 \pm 0.10$\\ 
         Demucs & $3.39 \pm 0.10$  & $2.55 \pm 0.10$ & $3.07 \pm 0.1$ & $2.85 \pm 0.11$ & $3.13 \pm 0.07$ & $3.21 \pm 0.07$ & $3.30 \pm 0.10$\\ 
         No Enhancement & $3.99 \pm 0.08$ & $2.48 \pm 0.11$  & $2.71 \pm 0.1$ & $3.04 \pm 0.12$ & $\mathbf{4.01 \pm 0.06}$ & $\mathbf{3.91 \pm 0.07}$ & $\mathbf{4.21 \pm 0.07}$\\ 

    \end{tabular}
    \caption{Mean Opinion Scores in a human listening test. Each column contains the ratings for a single perturbation type: EQ, additive background noise at different signal-to-noise ratios (SNR), and reverb at different direct-to-reverberant ratios (DRR).}
    \label{tab:ablation}
\end{table*}

The Mel2Mel generator described in Section \ref{ssec:conditional_image_synthesis} consists of 2 downsampling blocks, each containing a 2D convolutional kernel of size 3 and stride 2, instance normalization \cite{ulyanov2017instance} and ReLU activation functions. This is followed by 3 ResNet blocks \cite{7780459} with kernel size 3 and instance normalization. Finally, the representation is upsampled back to the original dimensionality of the input with two upsampling blocks, each containing a transposed convolutional kernel of size 3 and stride 2, instance normalization, and ReLU activation functions.
The Mel2Mel discriminator is a fully convolutional model made up of three blocks, each containing a convolutional kernel of size 4 and stride 2, instance normalization, and LeakyReLU activation function. The last layer does not have any normalization or activation function.
Both the generator and discriminator are trained with batch size of 64 and learning rate of 0.0002 for 200 epochs.

The Diffwave model described in Section \ref{ssec:music_vocoding} uses the architecture and training objective described in \cite{kong2021diffwave}. We train this model for 3000 epochs using a batch size of 8 and a learning rate of 0.0002. 

\subsection{Baselines}
We evaluate our approach against two separate baselines. 
First, we pair Mel2Mel for mel-spectrogram enhancement with inverse mel-scaling and the Griffin-Lim algorithm for musecoding. Both inverse mel-scaling and Griffin-Lim require solving optimization problems~\cite{yang2021torchaudio}, so we run both solvers for 100 iterations, which yields a per-sample runtime comparable to that of the Diffwave musecoder. 

Our second baseline is an end-to-end approach for music enhancement. Namely, we use the Demucs model architecture \cite{defossez2021music} and train it using the $\ell_1$ reconstruction loss on our dataset of low- and high-quality recording pairs. This matches the original training objective used for this architecture on the task of music source separation. We train this model for 360 epochs with batch size 64 and learning rate 0.0003. We find that after this number of epochs the validation loss plateaus.

\subsection{Evaluation Metrics}
To evaluate the results of different enhancement models we conducted a Mean Opinion Score (MOS) test with human listeners on Amazon Mechanical Turk (AMT). Additionally, we evaluate enhancement methods by computing the frequency-weighted segmental SNR (fwSSNR) \cite{Hu2008EvaluationOO}, multi-resolution spectrogram loss (MRS) \cite{yamamoto2020parallel}, $\ell_1$ spectrogram distance, and Fréchet Audio Distance (FAD) \cite{kilgour2019frechet} between enhanced and clean reference signals. In Section \ref{ssec:objective_metrics} we analyze the effectiveness of these objective metrics at approximating human listener ratings in the music domain.

\section{Results}
\label{sec:results}

\subsection{Mean Opinion Score Test}
\label{ssec:mos_test}
To evaluate our proposed Mel2Mel + Diffwave music enhancement model, we conducted an MOS test with human listeners on AMT. We used 200 audio samples from our test set, added 8 different types of simulated degradation, and passed these low-quality waveforms through our method, Mel2Mel + Griffin-Lim, and Demucs. The low-quality, enhanced, and ground truth high-quality samples were then presented to human listeners who were asked to give a quality score from 1 to 5. We used the ground truth high-quality recording as high anchor and the same recording with 0 dB white noise as low anchor. 

Each Human Intelligence Task (HIT) started with a screening test in which human listeners were required to identify which one of 5 audio samples sound the same as a reference sample. 4 out of the 5 samples are passed a small amount of effects including low pass filters, high pass filters, comb filters, and added noise. Passing the screening test was required to continue. The rest of the HIT consisted of 34 tests in which 4 were validation tests to check if listeners were paying attention. If they failed the validation test, the entire HIT was invalidated. 

In the end we collected 9,095 answers from 211 listeners. The results shown in Table \ref{tab:mos} suggests that Mel2Mel with a Diffwave musecoder achieves the highest MOS with a score near that of clean audio from the dataset. 

\subsection{Perturbation Ablation Study}
\label{ssec:ablation}
To gain insight into which perturbations are handled most effectively by each enhancement model, we perform an ablation study isolating each perturbation introduced in the low-quality signal generative process. Table \ref{tab:ablation} contains mean opinion scores for each enhancement model applied to signals with randomly sampled EQ, additive noise with signal-to-noise ratios (SNR) of 5, 10, and 15 dB, and reverb with direct-to-reverberant ratios (DRR) of 0, 3, and 6 dB. 

This ablation shows that the Mel2Mel + Diffwave model excels at removing noise even at SNR values as low as 5 dB and at undoing 4-band equalization simulating a non-flat microphone frequency response. Interestingly, none of the models tested perform dereverberation very well, and in fact degrade signals that contain no noise and only simulated reverb. This may be due to train-test mismatch, since all samples enhanced during training time contained some level of additive noise.

This ablation also lends insight into the types of perturbations that affect human listeners' perception of music. From the difference between the scores given to clean samples and non-enhanced samples, it is clear that additive noise impacts the listener's perception significantly while reverb is mostly ignored.

\subsection{Perceptual Alignment of Objective Metrics}
\label{ssec:objective_metrics}

The results of the MOS test also provide a mechanism to evaluate how well objective metrics for audio quality align with human perception in the music domain. We measure fwSSNR, MRS, FAD, and $\ell_1$ spectrogram distance on the same samples submitted for MOS evaluation. We then take the mean score across all samples with a given perturbation type (i.e. SNR 5, DRR 0, etc.) and perform a Spearman rank correlation with the mean scores measured in the human MOS test. In Table \ref{tab:objective_metrics} we show the rank correlation for each objective metric. We find that none of the four metrics evaluated correlate very strongly with human opinion scores, the highest achieving a rank correlation of 0.56. 

We also identify particular failure modes of these metrics. All four metrics fail to identify robotic artifacts induced by the Griffin-Lim algorithm and actually rate the Mel2Mel + Griffin-Lim model as the best of all models we tested. Additionally, fwSSNR, MRS, and $\ell_1$ spectrogram distance all fail to identify additive noise effectively, and rate non-enhanced samples at SNR values of 10 and 15 dB as being better than any enhancement model output. FAD does not have this failure mode.

\begin{table}[]
    \centering
    \begin{tabular}{c|c}
         Enhancement Metric & Rank Correlation with MOS \\ \hhline{-|-}
         fwSSNR & 0.5 \\ 
         $-$MRS &  0.56 \\ 
         $-$L1 & 0.4 \\ 
         $-$FAD & 0.53 \\ 
    \end{tabular}
    \caption{Spearman rank correlation between MOS test ratings and audio enhancement metrics.}
    \label{tab:objective_metrics}
\end{table}

\subsection{Alternate Training Schemes}
\label{ssec:training_schemes}
In Section \ref{ssec:modeling_approach} we motivated approaching music enhancement by training two decoupled models that separately handle mel-spectrogram enhancement and musecoding. Here, we investigate training schemes for these models other than independently training them on their respective tasks. In addition to independent training, we (1) finetune the Mel2Mel generator and Diffwave musecoder jointly using the Diffwave objective, (2) train the models sequentially by first training the musecoder and then training the Mel2Mel generator with musecoder parameters frozen, and (3) train the Mel2Mel generator and musecoder jointly as a single model using the Diffwave objective. 

Table \ref{tab:training_evaluation} shows the performance of the resulting models. In Section \ref{ssec:objective_metrics} we discussed the reliability of using these metrics for evaluating algorithms, and find that FAD is the most perceptually aligned metric when it comes to denoising. Given this observation, our results suggest that joint training may yield better denoising performance than independent training. Joint training has the added benefit that only a single model is trained using a non-adversarial objective.
However, this comes with the downside that the trained model cannot be split into enhancement and musecoding sub-models.
Future work could focus on further exploring such training schemes.

\begin{table}[] 
    \centering
    \begin{tabular}{c||c|c|c|c}
         Model & fwSSNR \textuparrow & MRS \textdownarrow & L1 \textdownarrow & FAD \textdownarrow\\ \hline
         Independent Training & \textbf{9.04} & \textbf{1.40} & \textbf{1.50} & 4.73\\ 
         Joint Fine-tuning & 7.61 & 1.57 & 1.57  & 4.54\\ 
         Joint Training & 6.58 & 1.65 & 1.69  & \textbf{3.98}\\ 
         Sequential Training & 8.23 & 1.80 & 1.83  & 5.54\\ 
         No Enhancement & 6.96 & 1.89 & 2.16  & 5.90\\ 

    \end{tabular}
    \caption{Performance of Mel2Mel + Diffwave enhancement models using different training schemes}
    \label{tab:training_evaluation}
\end{table}

\section{Conclusion}
\label{sec:conclusion}
We propose a music enhancement model that decomposes the task into mel-spectrogram enhancement and waveform synthesis from mel-spectrograms. 
This model was trained using high-quality samples from a public dataset paired with low-quality samples generated by simulating artifacts that typically appear in amateur recordings.
A human MOS test shows that this model outperforms state-of-the-art baselines. 
Additionally, we found that current objective metrics for audio enhancement do not accurately reflect human perception of music.
We hope this work encourages researchers to further advance the rather unexplored and yet timely topic of automatic music enhancement, either by designing more performant models or by proposing metrics that better align with human music perception.

\bibliographystyle{IEEEbib}
\bibliography{refs}
\end{document}